\documentclass[10pt,times,twocolumn]{article}

\usepackage{amssymb}
\usepackage{bibentry}
\usepackage[figuresright]{rotating}

\usepackage{authblk}
\usepackage{amsmath}
\usepackage{graphicx}

\usepackage{qcircuit}
\usepackage{algorithm}
\usepackage{algorithmic}
\usepackage{siunitx}
\usepackage{subcaption}
\usepackage{threeparttable}
\usepackage{url}
\usepackage{color}
\usepackage{setspace}
\usepackage{microtype}

\graphicspath{{./fig/}}

\newtheorem{msgm}{Minimum CNOT Gate Mapping (MCGM)}

\newcommand{\CX}{\ensuremath{\mathit{CX}}}
\newcommand{\SP}{\ensuremath{\mathit{SP}}}
\newcommand{\K}{K}

\begin{document}

\title{Optimization of Quantum Circuit Mapping using Gate Transformation and Commutation}

\author[1]{Toshinari Itoko\thanks{itoko@jp.ibm.com}}
\author[1]{Rudy Raymond}
\author[1]{Takashi Imamichi}
\author[1]{Atsushi Matsuo}
\affil[1]{IBM Research - Tokyo}

\date{}

\twocolumn[
  \begin{@twocolumnfalse}
\maketitle
\vspace{-5mm}
\begin{abstract}
This paper addresses quantum circuit mapping for Noisy Intermediate-Scale Quantum (NISQ) computers. 
Since NISQ computers constraint two-qubit operations on limited couplings, an input circuit must be transformed into an equivalent output circuit obeying the constraints.
The transformation often requires additional gates that can affect the accuracy of running the circuit.
Based upon a previous work of quantum circuit mapping that leverages gate commutation rules,
this paper shows algorithms that utilize both transformation and commutation rules.
Experiments on a standard benchmark dataset confirm the algorithms with more rules can find even better circuit mappings compared with the previously-known best algorithms.
\end{abstract}
\vspace{5mm}
  \end{@twocolumnfalse}
]

\section{Introduction} \label{sec:introduction}
Despite the limitation on the number of qubits, types of operations, and the operation time,
Noisy Intermediate-Scale Quantum (NISQ) computers are expected
to be useful in the near future for many applications of quantum chemistry, machine learning, optimization, and
sampling~\cite{preskill2018quantum}. To be able to run a quantum algorithm on NISQ devices for such applications,
the quantum algorithm is first transformed into a series of quantum operations (or quantum gates) on the quantum
bits (or qubits). A set of quantum gates that consists of arbitrary one-qubit rotation gates
and two-qubit controlled-NOT (or CNOT) gates is universal, i.e., any quantum operation can be realized by a combination
of such basic gates. However, because CNOT gates cannot be applied on all pairs of qubits of a NISQ device,
an input circuit must be transformed into an output circuit that obeys the limitation of the device.
The transformation can further limit the usability of the device, and therefore, a better compiler
is one of the most essential elements to utilize a NISQ device.

In this paper, we address the task on how to \emph{compile} a given quantum circuit
(i.e., transform it into an equivalent circuit) so that it can be run on NISQ devices.
We call such a task \emph{quantum circuit mapping} as it deals with mapping qubits
in the given quantum circuit to the actual qubits of NISQ devices with necessary additional gates
to obey the two-qubit operational constraints.

\begin{figure}[tb]
  \centering
  \begin{subfigure}[b]{0.2\textwidth}
    \[
    \Qcircuit @C=1em @R=.7em {
    \lstick{b_1} & \ctrl{1} & \qw      & \qw      & \targ    & \qw \\
    \lstick{b_2} & \targ    & \ctrl{1} & \gate{R_z} & \ctrl{-1}& \qw \\
    \lstick{b_3} & \ctrl{1} & \targ    & \qw      & \qw      & \qw \\
    \lstick{b_4} & \targ    & \qw      & \qw      & \qw      & \qw
    }
    \]
    \caption{Quantum circuit}\label{fig:qc}
  \end{subfigure}
  ~
  \begin{subfigure}[b]{0.2\textwidth}
      \includegraphics[clip, width=\textwidth]{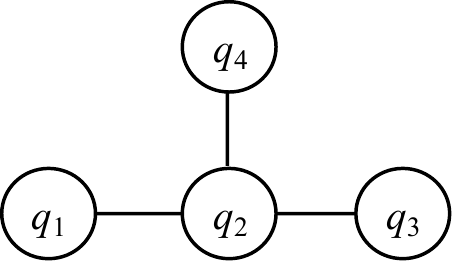}
      \caption{Coupling graph}\label{fig:cg}
  \end{subfigure}
  \caption{Quantum circuit (left), and coupling graph of a NISQ device that limits two-qubit operations (right).
  The problem is
  finding a mapping of $\{b_i\}$ to $\{q_j\}$ and additional gates so that the circuit can be run
  in the NISQ device with minimum cost.
  }
\end{figure}

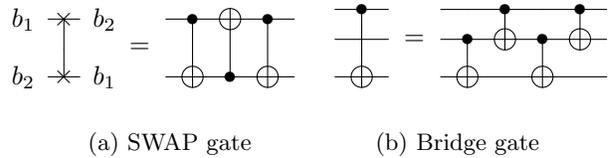
\begin{figure}[tb]
  \centering
	\begin{subfigure}[b]{.25\textwidth}
		\[
		\Qcircuit @C=.6em @R=.5em {
			\lstick{b_1} & \qswap      & \rstick{b_2} \qw & & & \ctrl{2} & \targ     & \ctrl{2} & \qw\\
			& \qwx        & & \push{\rule{1em}{0em}=} & & & & &\\
			\lstick{b_2} & \qswap \qwx & \rstick{b_1} \qw & & & \targ    & \ctrl{-2} & \targ    & \qw
		}
		\]
		\caption{SWAP gate}\label{fig:swap}
	\end{subfigure}
	\begin{subfigure}[b]{.20\textwidth}
		\[
		\Qcircuit @C=.6em @R=.6em {
			& \ctrl{2} & \qw &               & & \qw      & \ctrl{1} &  \qw      & \ctrl{1}& \qw \\
			& \qw      & \qw & \push{=} & & \ctrl{1} & \targ    & \ctrl{1} & \targ    & \qw  \\
			& \targ    & \qw &                & & \targ     & \qw      & \targ      & \qw     & \qw
		}
		\]
		\caption{Bridge gate}\label{fig:bridge}
	\end{subfigure}\quad
	\caption{Supplementary gates to be used in circuit mapping}
	\label{fig:supp}
\end{figure}

Figure~\ref{fig:qc} is an example
of an input circuit that uses a single-qubit gate, called the phase-shift gate (denoted as $R_z$),
and CNOT gates (depicted with vertical lines whose ends are $\bullet$ and $\oplus$ denoting
the control and target qubits, respectively). These gates act on qubits each represented by
a horizontal line in the figure, and are applied from left to right.
Figure~\ref{fig:cg} is an example of a \emph{coupling graph},
which is a graph representing which pairs of qubits can be coupled, i.e., be used by two-qubit CNOT gates.
Currently available NISQ devices, such as IBM Q systems~\cite{ibmq}, are limited by their coupling architecture
in the sense that not all pairs of qubits can be directly employed due to hardware issues.
If we map the qubit $b_i$ (as numbered in the circuit) in Fig.~\ref{fig:qc} to $q_i$
(as numbered in the hardware) for $i = 1,\ldots, 4$, the CNOT gate from $b_3$ to $b_4$ is not directly possible.
This can be resolved by swapping control and/or target qubits with their neighboring qubits until they are adjacent in the coupling graph.
Such swapping can be carried out by so-called SWAP gates (Fig.~\ref{fig:swap}).
Another way to run a CNOT gate between non-adjacent qubits is replacing it with a sequence of CNOT gates.
This composite gate is called a Bridge gate and is composed of four CNOT gates as shown in Fig.~\ref{fig:bridge}.
In both cases, we need to add supplementary gates to the original circuit, resulting in more noise and computational
time since CNOT gates are noisy and take significantly more time for computation than single-qubit gates.
Thus, minimizing additional gates in the circuit mapping is important as it translates into minimizing the increase
of time and noise cost to run the circuit. Indeed, there has been a large body of work addressing the quantum circuit mapping
(see Section~\ref{sec:motivation} for details). However, most of those circuit mappings are not optimal because they add
supplementary gates layer by layer in the given quantum circuit.

In our recent paper~\cite{itoko2019quantum}, we discussed the drawback of resolving the two-qubit operations layer by layer in many existing approaches
and showed the importance of taking into account gate commutation rules. We demonstrated
how four simple commutation rules, which consists of commuting CNOT gates and/or single-qubit gates, can lead to quantum circuit
mapping with significantly fewer additional SWAP gates.

In this paper, we
further extend the formulation with commutation rules and transformation rules of running CNOT gates on non-adjacent qubits
by choice of replacing with Bridge gates or inserting SWAP gates.
Although SWAP and Bridge gates seem to require different numbers of CNOT gates (three as in Fig.~\ref{fig:swap}, and four as in Fig.~\ref{fig:bridge}), a
Bridge gate results in running the CNOT gate, whereas a SWAP gate only swaps its two qubits. Thus, the total number of additional CNOT gates to run a CNOT gate
between an unconnected qubit pair is the same for both SWAP and Bridge gates. However, the SWAP gate permutes the ordering of the logical qubits,
whereas the Bridge gate does not. Depending on the sequence of gates afterwards, the selection of SWAP and Bridge gates can affect the possibility of
further reducing the additional CNOT gates in the mapping.

Experiments on a well-known benchmark dataset show that the new transformation rules can lead to even better cost of mapping
than with known approaches: 14.2\% fewer additional CNOT gates on average than those without the new rules.
Hence, this paper demonstrates the generality and the high potential of the formulation of quantum circuit mapping
that takes into account transformation and commutation rules for quantum gates.

\section{Motivation and Related Work} \label{sec:motivation}
The main contribution of this paper is proposing algorithms using both gate commutation and transformation rules (i.e., the Bridge and SWAP gates)
that result in better circuit mappings for existing benchmark circuits. There have been many studies of quantum circuit mapping, e.g., those dealing
with the circuit mapping on 1D-chain (known as linear nearest neighbor (LNN)) topology
\cite{matsuo2011changing,hirata2011efficient,saeedi2011synthesis,chakrabarti2011linear,shafaei2013optimization,wille2014optimal,rahman2015synthesis},
those on 2D-grid nearest neighbor topology \cite{shafaei2014qubit,lye2015determining,wille2016look,ruffinelli2017linear}, and, like ours, those on the general topology
of NISQ devices \cite{zulehner2018efficient,bhattacharjee2017depth,siraichi2018qubit,venturelli2017temporal,li2019tackling,cowtan2019qubit,wille2019mapping}.
A recent study \cite{paler2018nisqcompile}
discussed the complexity of circuit compilers on NISQ devices and their search-space structure. However, to the best of our knowledge,
those studies either did not fully consider the gate commutation rules or did not consider transformation rules other than SWAP gates.

\begin{figure}[tb]
	\centering
	\begin{minipage}{0.45\textwidth}
\[
\Qcircuit @C=1em @R=.7em {
  & \push{\mbox{layer 1}} & \push{\mbox{layer 2}} &  & \push{\mbox{layer 3}} & \\
  \lstick{b_1} & \ctrl{1}_{\hspace{9mm}\CX_1} & \qw   & \qw     & \targ_{\hspace{12mm}\CX_4} & \qw \\
  \lstick{b_2} & \targ & \ctrl{1}_{\hspace{13mm}\CX_3} & \gate{R_z} & \ctrl{-1} & \qw \\
  \lstick{b_3} & \ctrl{1}_{\hspace{9mm}\CX_2} & \targ & \qw    & \qw & \qw \\
  \lstick{b_4} & \targ & \push{\rule{0em}{.7em}} \qw & \qw & \qw & \qw
\gategroup{1}{2}{5}{2}{.5em}{--}
\gategroup{1}{3}{5}{3}{.5em}{--}
\gategroup{1}{5}{5}{5}{.5em}{--}
}
\]
	\end{minipage}
\\
	\begin{minipage}{0.45\textwidth}
	\begin{subfigure}[b]{0.55\textwidth}
		\includegraphics[width=1.0\textwidth]{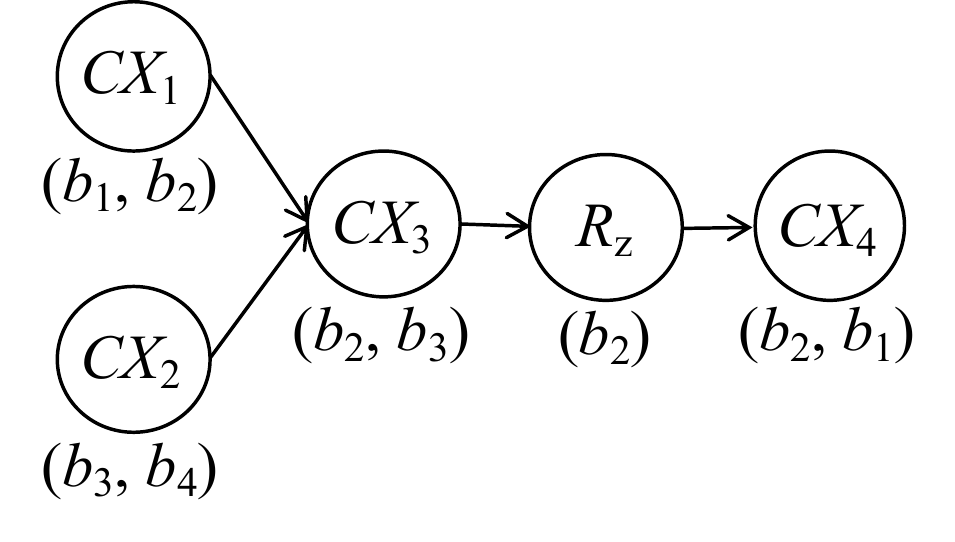}
		\caption{Standard DAG}\label{fig:standard}
	\end{subfigure}
~
	\begin{subfigure}[b]{0.40\textwidth}
		\includegraphics[width=1.0\textwidth]{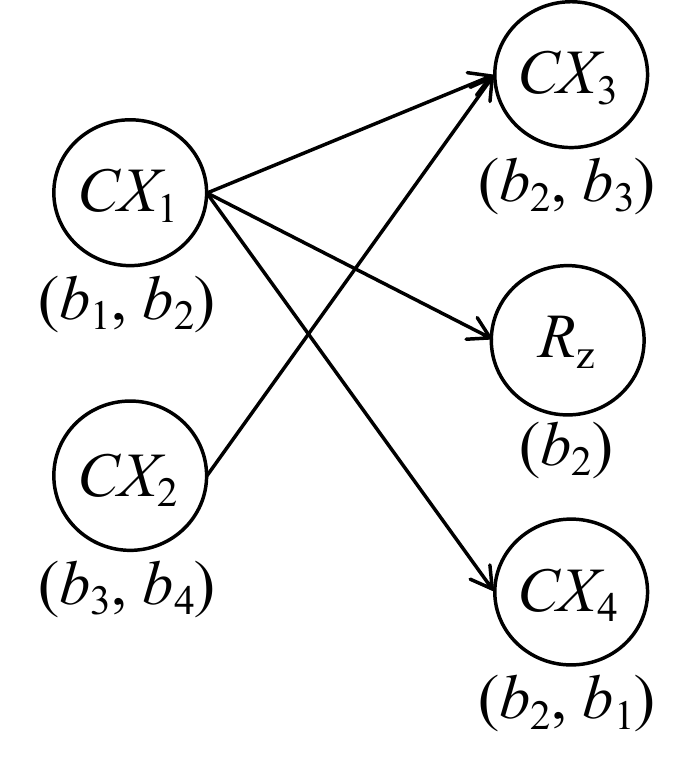}
		\caption{Extended DAG}\label{fig:extended}
	\end{subfigure}
	\end{minipage}
	\caption{Quantum circuit with $3$ layers (top) and its dependency graphs (bottom)}
	\label{fig:layers}
\end{figure}

With regard to gate commutation rules, many previous studies assumed a \emph{fixed-layer formulation},
where the layers of a quantum circuit are given as input and all gates in a layer must be mapped, and when necessary
resolved with SWAP gates, before any gate in the next layer. A \emph{layer} is defined as a set of CNOT gates that can be executed
in parallel (see the top part of Fig.~\ref{fig:layers}). Previously, we pointed out that the fixed-layer formulation may lead to suboptimal solutions, and proposed a formulation of circuit mapping that uses the commutation
rules to avoid suboptimal solutions~\cite{itoko2019quantum}.

\begin{figure}[tb]
	\centering
	\begin{subfigure}[b]{0.25\textwidth}
		\[
		\Qcircuit @C=.7em @R=.3em {
			& \gate{R_z} & \ctrl{1}  & \qw &          & & \ctrl{1}  & \gate{R_z} & \qw \\
			&            &           &     & \push{=} & &           &            & \\
			& \qw        & \targ\qwx & \qw &          & & \targ\qwx & \qw        & \qw
		}
		\]
		\caption{$R_z$--control}\label{fig:rule-a}
	\end{subfigure}
	~
	\begin{subfigure}[b]{0.2\textwidth}
		\[
		\Qcircuit @C=.7em @R=.5em {
			& \ctrl{2} & \ctrl{1} & \qw & \push{\rule{0em}{1em}} & & \ctrl{1} & \ctrl{2} & \qw \\
			& \qw      & \targ    & \qw & \push{=}               & & \targ    & \qw      & \qw \\
			& \targ    & \qw      & \qw &                        & & \qw      & \targ    & \qw
		}
		\]
		\caption{Control--control}\label{fig:rule-b}
	\end{subfigure}
	~
	\begin{subfigure}[b]{0.25\textwidth}
		\[
		\Qcircuit @C=.7em @R=.3em {
			& \qw      & \ctrl{1}  & \qw &          & & \ctrl{1}  & \qw      & \qw \\
			&          &           &     & \push{=} & &           &          & \\
			& \gate{R_x} & \targ\qwx & \qw &          & & \targ\qwx & \gate{R_x} & \qw
		}
		\]
		\caption{$R_x$--target}\label{fig:rule-c}
	\end{subfigure}
	~
	\begin{subfigure}[b]{0.2\textwidth}
		\[
		\Qcircuit @C=.7em @R=.5em {
			& \ctrl{2} & \qw      & \qw & \push{\rule{0em}{1em}} & & \qw      & \ctrl{2} & \qw \\
			& \qw      & \ctrl{1} & \qw & \push{=}               & & \ctrl{1} & \qw      & \qw \\
			& \targ    & \targ    & \qw & \push{\rule{0em}{1em}} & & \targ    & \targ    & \qw
		}
		\]
		\caption{Target--target}\label{fig:rule-d}
	\end{subfigure}
	\caption{Standard commutation rules of quantum gates}
	\label{fig:rules}
\end{figure}
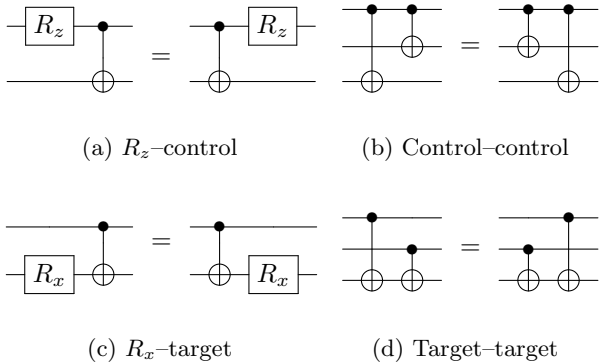

The drawback of the fixed-layer approach can be seen from the circuit in Fig.~\ref{fig:qc} when the coupling graph
is as shown in Fig.~\ref{fig:cg}. The number of additional SWAP gates by the fixed-layer formulation is two, but it is one if we allow
to be changed the order of gates by applying the standard commutation rules of quantum gates illustrated in Fig.~\ref{fig:rules}.
The reason is as follows.

For the circuit in Fig.~\ref{fig:layers}, by the $R_z$--control rule (Fig.~\ref{fig:rule-a}), we can move the CNOT gate $\CX_4$
to the left so that it precedes $R_z$. Now, the CNOT gates $\CX_3$ and $\CX_4$ share
the control qubit, and by the control--control rule (Fig.~\ref{fig:rule-b})
we can move $\CX_4$ to the left. As a result, when $b_i$ is mapped to $q_i$,
$\CX_1$ and $\CX_4$ can be applied before swapping $b_2$ and $b_3$ to run
$\CX_2$, $\CX_3$, and $\CX_5$. This is not possible for the circuit {\emph{with fixed layers}} in Fig.~\ref{fig:layers}
because $\CX_4$ in layer 3 cannot precede $\CX_2$ in layer 2.

Because the partial order of gates in a circuit is naturally modeled with a directed acyclic graph (DAG),
some studies \cite{li2019tackling,qiskit} have considered trivial commutation between consecutive gates that do not share qubits. For example, Fig.~\ref{fig:standard}
is a DAG representation of the circuit in the top part of Fig.~\ref{fig:layers}. We call this \emph{the standard-DAG formulation}.
Although it is more flexible than the fixed-layer one, it still may fall into suboptimal solutions: it leads to two additional SWAP gates
for the circuit in Fig.~\ref{fig:layers}.

To overcome the suboptimality problem, Itoko et al.~\cite{itoko2019quantum}
proposed a novel formulation taking into account gate commutation rules in combination with the concept \emph{dependency graph} that extends the standard-DAG.
The dependency graph depicts the true partial order of gates as imposed by the commutation rules.
For example, the dependency graph of the circuit in Fig.~\ref{fig:layers} is shown in Fig.~\ref{fig:extended}.
The dependency graphs with fewer commutation rules than those in the work of Itoko et al.~\cite{itoko2019quantum} were first considered by Matsuo et al.~\cite{matsuo2011changing}.
Venturelli et al.~\cite{venturelli2017temporal} considered commutation rules specific to their circuits of interest and proposed a solution based on a temporal planner;
however, they did not provide systematic methods exploiting the rules, unlike our approaches.
Recently, Li et al.~\cite{li2019tackling} independently proposed a heuristic algorithm very similar to ours.
To be precise, although they do not explicitly consider commutation rules in their heuristic,
it is easy to modify their algorithm to take into account our proposed dependency graph.
However, like other mappings, they only use SWAP gates to map two-qubit gates to NISQ devices.

In this paper, we consider Bridge gate as well as SWAP gate as a transformation rule used in mapping.
There are only a few studies considering Bridge gate
(aka, chain template), e.g.~\cite{siraichi2018qubit,shende2006synthesis}.
Although their main purpose was for the synthesis of quantum circuit from a unitary matrix, Shende et al.~\cite{shende2006synthesis} hinted that quantum circuit mapping can be solved with only chain templates.
Siraichi et al.~\cite{siraichi2018qubit} provided exact and heuristic algorithms using both SWAP and Bridge gates, but they reported the implementation of their heuristic in Qiskit~\cite{qiskit} did not perform well
for non-synthetic circuits, which may be caused by not using the dependency graph.

\begin{figure*}[t]
	\centering
	\includegraphics[width=1.0\textwidth]{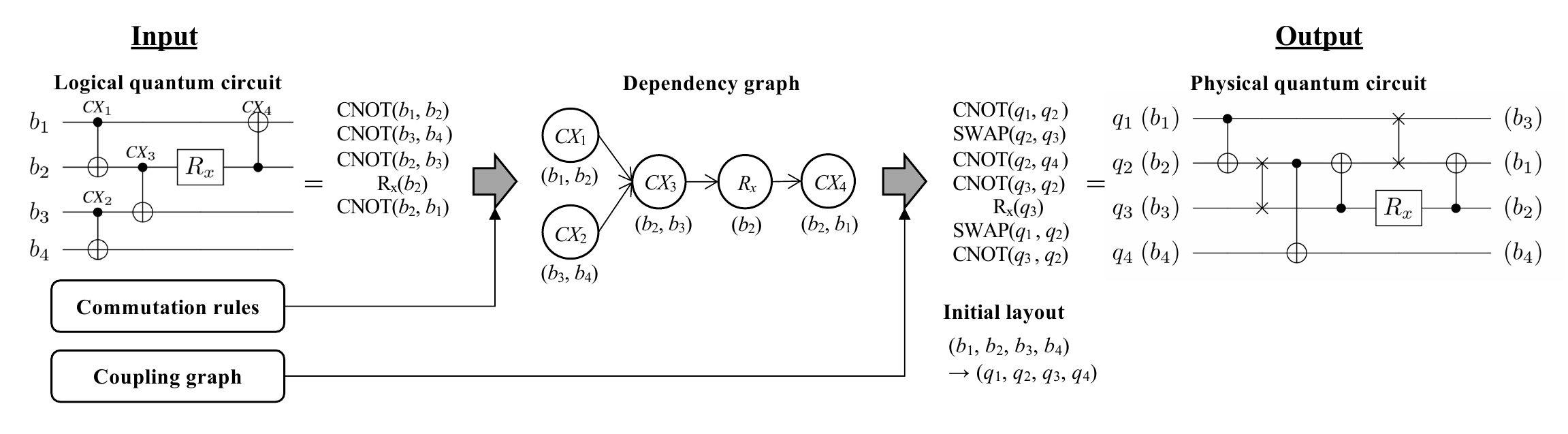}
	\caption{Overview of our circuit mapping based on Minimum CNOT Gate Mapping (MCGM) formulation}
	\label{fig:whole}
\end{figure*}

\section{Problem Formulation} \label{sec:form}

We propose a formulation of quantum circuit mapping that takes into account gate
commutation and transformation rules to minimize the number of additional CNOT gates.
We first describe the notations used hereafter.

We use the term \emph{logical qubits} to represent the qubits
in the original quantum circuit (denoted by $B$).
This is not to be confused with the term logical
qubits in the context of quantum error correction.
We use the term \emph{physical qubits} to represent
the qubits of the quantum hardware (denoted by $Q$).
For example, $B = \{b_i\}$ and $Q = \{q_j\}$ denote logical qubits and physical qubits
in Fig.~\ref{fig:qc} and~\ref{fig:cg}, respectively.
Similarly, we call the input circuit the \emph{logical (quantum) circuit} and
its corresponding output of quantum circuit mapping the \emph{physical (quantum) circuit}.
The physical circuit is equivalent to its logical circuit in terms of computation but may
contain additional gates due to coupling constraints. A coupling graph $C = (V_C, E_C)$
is a graph with physical qubits as nodes $V_C (=Q)$ and coupled physical qubits as edges $E_C$.
We assume that $|B| = |Q|$ by adding ancillary qubits to $B$ if $|B| < |Q|$.

A mapping of logical qubits to physical qubits can be seen as a permutation function $B \to Q$,
which we call a \emph{layout}. A logical/physical circuit can be seen as a list of
gates acting on logical/physical qubits.
We say a physical circuit $\hat{L}$ \emph{complies} with a coupling graph $C$
if, for any CNOT gate in $\hat{L}$, its control and target qubits are adjacent in $C$.
A mapping of a logical circuit $L$ to a quantum computer with a coupling graph $C$
can be represented by an \emph{initial layout} $l_0$ of qubits
and a physical circuit $\hat{L}$ that complies with $C$.

The objective of quantum circuit mapping is to find a cost-optimal physical circuit of
a logical circuit. In this paper, we consider the number of additional CNOT gates as the cost of mapping
because they are usually much noisier and slower to run than single-qubit gates.

The problem of finding a corresponding physical circuit of a logical circuit with
the minimum number of CNOT gates is formalized as follows.
\bigskip
\begin{msgm}
	Given a logical circuit $L$, the coupling graph $C$ of a quantum hardware, and
	a set of commutation rules $\mathcal{R}$,
	find an initial layout $l_0$ and an equivalent physical circuit $\hat{L}$ with the fewest CNOT gates that complies with $C$.
\end{msgm}

Note that fixed-layer and standard-DAG formulations in many previous work can be regarded
as special cases of MCGM: the former with $\mathcal{R} = \emptyset$, and the latter with $\mathcal{R}$
containing trivial commutation between consecutive gates that do not share qubits. In the next section, we propose
algorithms that utilize the commutation rules in Fig.~\ref{fig:rules} for $\mathcal{R}$ in MCGM because they are complete for
rules involving two consecutive gates from the universal gate set \{$R_x$, $R_z$, CNOT\}, and the transformation rules of CNOT gates
with SWAP or Bridge gates. Here $R_x$ and $R_z$
denote single-qubit rotation (respectively, around the $x$-axis and $z$-axis) gates. Without loss of generality, we assume
any input circuit is composed of gates from the universal gate set.


\section{Algorithms} \label{sec:algorithm}

We show how to solve the Minimum CNOT Gate Mapping (MCGM) with an exact algorithm based on dynamic programming (DP),
and a heuristic algorithm based on a look-ahead scheme.
To minimize the number of additional CNOT gates,
both algorithms resort to minimizing the number of additional SWAP and Bridge gates,
each of which requires three additional CNOT gates.
Note that a Bridge gate can only be applied to run a CNOT gate
when the distance between control and target qubits of the CNOT gate is exactly two.
Both algorithms are extensions of algorithms minimizing SWAP gates introduced
in~\cite{itoko2019quantum}. The exact algorithm can obtain optimal solutions
but can only be used for small circuits because of its long computational time.
The heuristic algorithm can obtain good solutions for a larger circuit within a reasonable amount
of computational time. The building blocks of the algorithms are dependency graphs
and blocking gates, as explained below. Figure \ref{fig:whole} shows an overview
of our circuit mapping algorithms.

\paragraph{Dependency Graph}
A dependency graph $D=(V_D, E_D)$ is a DAG that represents the precedence
relation of the gates in an input circuit taking commutation rules into account.
The nodes $V_D$ correspond to the gates in the logical circuit, and the edges $E_D$ to
the dependencies in the order of gates. A gate $g_i$ must precede $g_j$ under the commutation
rules if and only if there exists a path from $g_i$ to $g_j$ on $D$.
See Fig.~\ref{fig:layers} for examples of dependency graphs: one for trivial commutation
rules of gates that do not share qubits, and the other for commutation rules in Fig.~\ref{fig:rules}.
The labels below a gate (node) in the dependency graphs represent
the logical qubits that the gate acts on, e.g., $\CX_1$ acts on $(b_1, b_2)$.


We can construct a dependency graph $D$ from a given logical quantum circuit, as in Fig.~\ref{fig:qc},
straightforwardly as follows (see \ref{sec:const-dg} for the details).
We check all pairs of the gates $V_D$ in the circuit, and include an edge $(g_i, g_j)$ in $E_D$
if the following conditions are satisfied:
(1) $g_i$ and $g_j$ share at least one logical qubit $b$, and
(2) the set of gate symbols from $g_i$ to $g_j$ on each shared $b$ in the logical quantum circuit is not a subset of ${R_z, \bullet}$ (i.e, $R_z$ gate and control of CNOT gate) or a subset of ${R_x, \oplus}$ (i.e., $R_x$ gate and target of CNOT gate).

\begin{figure}
	\centering
	\includegraphics[width=0.4\textwidth]{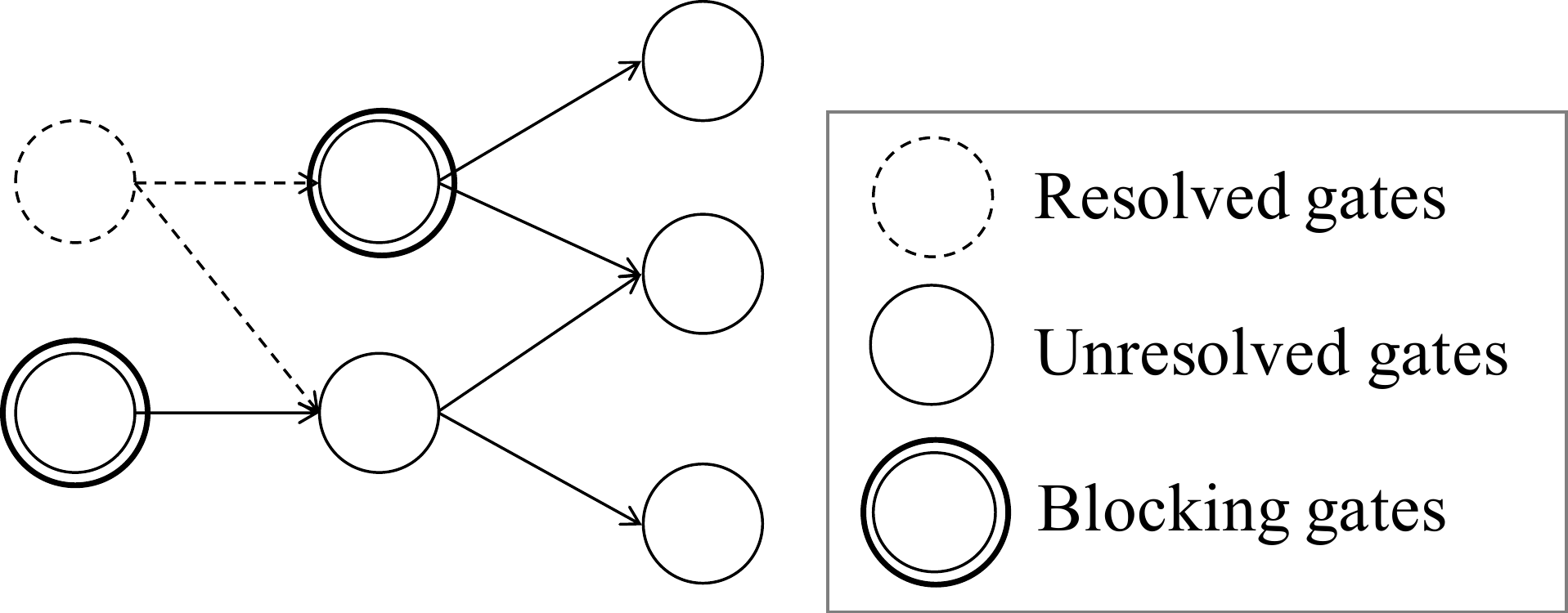}
	\caption{Blocking gates in dependency graph}
	\label{fig:blocking}
\end{figure}

\paragraph{Blocking Gates}
Both exact and heuristic algorithms take the same strategy by resolving
the gates in the dependency graph $D$ one by one.
They maintain the progress by recording \emph{blocking gates},
which are defined as leading unresolved gates in $D$ as shown in Fig.~\ref{fig:blocking}.
For given gates $G$ (usually blocking gates),
we define \emph{blocking gates from $G$ for $l$}  by blocking gates after running leading gates
that comply with the coupling graph $C$ repeatedly starting from $G$ under the layout $l$.
At the beginning of mapping, the gates with no in-edges in $D$, say $G_0$,
can be seen as blocking gates. For a given initial layout $l$,
we define the \emph{initial blocking gates for $l$} (denoted by $K_0(l)$)
to be the blocking gates from $G_0$ for $l$.
%

\subsection{Exact algorithm} \label{sec:exact}
We design an exact algorithm to solve MCGM based on DP.
To achieve this, we define a \emph{state} similar to DP by $(l, \K)$, a pair of a layout $l$ and blocking gates $\K$.
%

A transition between a pair of states occurs when a SWAP or Bridge gate is inserted.
For example, let us consider the logical circuit and dependency graph as in Fig.~\ref{fig:whole}
and the coupling graph as in Fig.~\ref{fig:cg}. Then, the blocking gates from $\{\CX_1, \CX_2\}$ for an initial layout
$l_0: (b_1, b_2, b_3, b_4) \mapsto (q_1, q_2, q_3, q_4)$ is $\{\CX_2\}$.
This is because the leading $\CX_1$ can be run, but $\CX_2$ blocks the rest of the gates.
If we choose the SWAP gate between $b_2$ and $b_3$, SWAP($b_2, b_3$), (or equivalently SWAP($q_2, q_3$) under $l_0$),
it changes the state $\left(l_0, \{\CX_2\}\right)$
to another state $\left((b_1, b_2, b_3, b_4) \mapsto (q_1, q_3, q_2, q_4), \{\CX_4\}\right)$.
Alternately, we can choose the Bridge gate from $q_3=l_0(b_3)$ to $q_4=l_0(b_4)$ via $q_2$ to run $\CX_2 = \mathrm{CNOT}(b_3, b_4)$,
where for simplicity we denote the mapping of a logical qubit $b_i$ to a physical qubit $q_j$ under layout $l_0$ as $q_j = l_0(b_i)$.
The transformation of CNOT into a Bridge gate changes the state $\left(l_0, \{\CX_2\}\right)$ to another state $\left(l_0, \emptyset\right)$.
That is, the transformation does not change the layout but updates the blocking gates $\{\CX_2\}$ to $\emptyset$.
After $\CX_2$ is run as the Bridge gate, all the remaining gates can be run under the same layout $l_0$.
This example shows how Bridge gates are important to realize better mapping algorithms.

\begin{algorithm}[t]
		\small
	\caption{DP-based exact algorithm for MCGM}
	\label{algo:exact}
	\begin{algorithmic}[1]
		\REQUIRE A dependency graph $D = (V_D, E_D)$ with logical qubits $B$, a coupling graph $C = (V_C, E_C)$
		with physical qubits $Q$
		\ENSURE The minimum number of SWAP and Bridge gates in mapping of $D$ to $C$
		\STATE $S \leftarrow \{(l, \K_0(l)) \mid$ layout $l$ from $B$ to $Q\}$\quad\COMMENT{active states}
		\STATE $f(l, \K) \leftarrow 0$ for all initial state $(l, \K) \in S$
		\WHILE{True}
		\STATE $S' \leftarrow \emptyset$\quad\COMMENT{next active states}
		\FORALL{state $(l, \K)$ in $S$} \label{line:exa-out}
		\FORALL{edge $e$ in $E_C$} \label{line:exa-inn}
		\STATE $l' \leftarrow$ layout after swapping $e$ of $l$
		\STATE $\K' \leftarrow$ blocking gates from $\K$ for $l'$
		\IF{$f(l', \K')$ is not yet defined} \label{line:same-begin}
		\STATE $f(l', \K') \leftarrow 1 + f(l, \K)$ \label{line:update}
		\STATE $S' \leftarrow S' \cup \{(l', \K')\}$
		\ENDIF
		\STATE \textbf{if} $\K' = \emptyset$ \textbf{then return} $f(l', \K')$ \label{line:same-end}
		\ENDFOR
		\FORALL{CNOT gate $g$ (acting on $(b_i, b_j)$) in $K$}
		\IF{distance between $l(b_i)$ and $l(b_j)$ on $C$ = $2$}
		\STATE $l' \leftarrow l$\quad\COMMENT{no change in layout}
		\STATE $\K' \leftarrow$ blocking gates from $\K \setminus \{g\}$ for $l$  \\
		\STATE Do the same procedure between line \ref{line:same-begin} and \ref{line:same-end}
		\ENDIF
		\ENDFOR
		\ENDFOR
		\STATE $S \leftarrow S'$
		\ENDWHILE
	\end{algorithmic}
\end{algorithm}

Let $f(l, \K)$ denote the minimum number of additional SWAP and Bridge gates required
to reach a state $(l, \K)$ starting from any initial state.
Here an initial state is a state defined by $(l, \K_0(l))$ for an initial layout $l$.
The minimum number of SWAP and Bridge gates in a mapping is thus
the minimum value of $f(l, \emptyset)$ for all possible $l$s.
This can be computed as follows.
The algorithm first sets $0$ to $f$ of the initial states then checks the states that
can be reached with one SWAP or Bridge gate (and set $1$ to $f$s of the states that have not yet seen),
those reached with two SWAP and/or Bridge gates (and set $2$ to the newly checked states), and so on.
The algorithm terminates when it reaches a state whose set of blocking gates is empty.
Note that each state is activated at most once because of the minimality of $f$.
See Algorithm~\ref{algo:exact} for details. The optimal gate list $\hat{L}$ can be obtained
by recording the gates resolved by each of the swaps that update $f$ at line~\ref{line:update}
and traversing from the last state in the opposite direction.

\subsection{Heuristic algorithm} \label{sec:heuristic}
We first review the original heuristic algorithm~\cite{itoko2019quantum}
and then describe how to extend it to handle both SWAP and Bridge gates.

The original \emph{look-ahead} heuristic algorithm
takes into account not only the blocking gates but also the other unresolved gates
in the selection of a qubit pair to be swapped. The look-ahead algorithm is different
from those based on the fixed-layer formulation~\cite{wille2016look,zulehner2018efficient} in the sense that
it does not require any definition of layers.
The idea of the look-ahead mechanism is as follows.
With regards to a layout $l$, for each unresolved two-qubit gate $g$
we can compute the length of the shortest path between the two qubits of $g$ as
the minimal number of SWAP gates required to apply $g$.
Thus, the number of SWAP gates of the layout $l$ is proportional to the total
length of the shortest paths of all unresolved two-qubit gates.
Slightly modifying the layout $l$ with swapping $(q_i,q_j)$, denoted as $\tilde{l}(q_i, q_j)$,
may decrease the length of the shortest paths at some unresolved two-qubit gates;
but it may also increase those at other unresolved gates.
The look-ahead heuristic prefers modifying the layout so that the total length of the shortest paths is reduced.

The pseudocode of the look-ahead heuristic algorithm is shown in Algorithm~\ref{algo:heuristic}, where
the lines from \ref{line:bridge-begin} to \ref{line:bridge-end} are for considering Bridge gates.
Like the exact one, the heuristic algorithm also maintains a layout $l$ and blocking gates $\K$;
however, unlike the exact one, we assume that an initial layout is given.
\begin{algorithm}[tb]
		\small
	\caption{Look-ahead heuristic algorithm for MCGM}
	\label{algo:heuristic}
	\begin{algorithmic}[1]
		\STATE $C$: a coupling graph, $D$: a dependency graph
		\STATE Initialize blocking gates $\K$ as the gates with no in-edges in $D$ and layout $l$ as the given initial layout.
		\LOOP
		\STATE Run the gates complying with $C$ and update $\K$.
		\STATE \textbf{if} $\K = \emptyset$ \textbf{then terminate}
		\STATE Compute the swap score for each edge in $C$.
		\STATE Let $(q_s, q_t)$ be the edge with the highest swap score.
		\STATE Let $(b_{s'}, b_{t'}) \mapsto (q_s, q_t)$ be the current mapping in $l$.
		\STATE Let $S \subseteq K$ be CNOT gates whose acting qubits has distance two in $C$ for $l$. \label{line:bridge-begin}
		\IF{$S$ is not empty and the highest swap score $< 1$}
		\STATE Transform some $g \in S$ acting on $(q_s, q_t)$ in $l$ into Bridge gate between $(q_s, q_t)$.
		\STATE \textbf{continue}
		\ENDIF \label{line:bridge-end}
		\IF{the swap of the mapping at $(q_s, q_t)$ in $l$ decreases $\sum_{g \in \K}{ \SP(g, l)}$}
		\STATE Swap the mapping, i.e., $(b_{s'}, b_{t'}) \mapsto (q_t, q_s)$. \label{line:swap}
		\ELSE
		\STATE Choose any gate $\hat{g}$ in $\K$ and add swap gates that strictly decrease $c(l, \{\hat{g}\})$  until $\hat{g}$ is resolved. \label{line:special}
		\ENDIF
		\ENDLOOP
	\end{algorithmic}
\end{algorithm}
The heuristic algorithm updates $\K$ at the beginning of each loop.
If $\K$ is empty, it terminates. Otherwise, it selects a qubit pair to be swapped on the basis of its swap \emph{score},
i.e., the difference between the sum of the length of the shortest paths before and after swapping the qubit pair.
Note that the selected qubit pair should be an edge of the coupling graph $C$.
Let $U$ be a subset of unresolved CNOT gates including the current $\K$ and
$\SP(g, l)$ be the length of the shortest path from the two qubits of $g$ with regards to the layout $l$.
The (swap) score of a qubit pair $(q_i, q_j) \in E_C$ with $l$ and $U$ is defined as follows:
\begin{align*}
\textit{score}((q_i, q_j), l, U) &= c(l, U) - c(\tilde{l}(q_i, q_j), U),\\
c(l, U) &= \sum_{g \in U}{\gamma(g)\, \SP(g, l)},
\end{align*}
where $\tilde{l}(q_i, q_j)$ denotes the layout after swapping $(q_i, q_j)$ from $l$ and
$\gamma(g)$ denotes the weight of the shortest path of the gate $g$.
To prioritize swapping
a qubit pair reducing shortest paths of blocking gates and other unresolved gates close to them,
we let $\gamma(g)=1$ for $g \in \K$ and $\gamma(g) = \alpha^{d(\K, g)}$ ($0 < \alpha < 1$) for $g \notin \K$,
where $d(\K, g)$ denotes the longest path length among the paths from $g' \in \K$ to $g$.
\begin{table*}
	\centering
		\small
	\caption{Comparison of averages of optimal numbers of additional SWAP and Bridge gates of our formulation (in the Proposed column)
  and those of fixed-layer and standard-DAG formulation for two sets of 10 random circuits with five or six qubits.
  The average numbers of Bridge gates are stated in $(\cdot)$. The No-Bridge column lists
  the optimal numbers of SWAP gates from the original formulation~\cite{itoko2019quantum}.
  ($|B|$: Number of qubits in circuit)}
	\label{tab:vs-layer}
	\begin{tabular}{cc|rrrr} \hline
$|B|$	&	Coupling	&	Fixed-layer	&	Std-DAG	&	Proposed	&	No-Bridge	\\
 \hline
5	&	ibmqx4	&	9.4 (3.4)	&	8.9 (2.6)	&	7.2	 (1.1)	&	7.4	\\
5	&	LNN	   &	22.1 (6.3)	&	21.8 (7.0)	&	19.5 (4.4)	&	21.2	\\
6	&	$2 \times 3$	   &	11.8 (1.7)	&	11.7 (2.0)	&	10.1 (1.4)	&	10.7	\\
6	&	LNN	&	28.2 (6.3)	&	27.5 (8.0)	&	23.7 (5.3)	&	26.4	\\
 \hline
	\end{tabular}
\end{table*}
At each loop of Algorithm~\ref{algo:heuristic}, SWAP gates are added at either line~\ref{line:swap} or at line~\ref{line:special}.
The former always decreases the length of the shortest paths of blocking gates, while the latter always decreases the size of $\K$.
This guarantees that the algorithm terminates after a finite number of loops.

The algorithm can be extended to handle Bridge gates by defining the score for replacing a non-adjacent CNOT with a Bridge gate.
Since a Bridge gate does not change the layout, we can define
$\textit{score}((q_i, q_j), l, U) = 1$ for any $(q_i, q_j)$ such that
$(q_i, q_j) = (l(b_{i'}), l(b_{j'}))$ for a CNOT gate acting on $(b_{i'}, b_{j'})$
and the distance between $(q_i, q_j)$ is two. Such extension are depicted at
the lines from \ref{line:bridge-begin} to \ref{line:bridge-end} in Algorithm~\ref{algo:heuristic}.

\section{Experimental Results} \label{sec:experiment}

We conducted two sets of experiments: comparing the effectiveness of our formulation
against that of formulations with fewer commutation rules and one without Bridge gates,
and evaluating the performance of our heuristic algorithm
against those of state-of-the-art algorithms. Both were conducted on a laptop PC with
an Intel Core i7-6820HQ 2.7 GHz and 16 GB memory.

\subsection{Comparison to formulations with fewer commutation rules and one without Bridge gates} \label{sec:exp-exact}

In the first set of experiments, we compared the optimal numbers of additional gates with
our formulation to those with the other formulations; two formulations with fewer commutation rules, i.e. the fixed-layer and standard-DAG, and one without Bridge gates.
We obtained the optimal numbers of additional gates by applying our exact algorithm discussed in Section~\ref{sec:algorithm}--\ref{sec:exact}.
To obtain the optimal numbers for the formulations  with fewer commutation rules, we added extra edges to the dependency graphs
so that they represent the fixed layers and standard-DAG, and then we applied our exact algorithm to them.
We used two sets of 10 random circuits with five or six qubits. Each circuit contained 100 gates
in which each gate was either a $R_z$, $H$, or CNOT gate with probability of 25\%, 25\%, or 50\%, respectively.
Here $H$ is a single-qubit gate called the Hadamard gate.

Table~\ref{tab:vs-layer} compares additional gates for our formulation and the other formulations.
Recall that each of SWAP and Bridge gate introduces three extra CNOT gates as in Fig.~\ref{fig:swap} and \ref{fig:bridge},
so the total number of additional CNOT gates is exactly three times the number of additional SWAP and Bridge gates.
By comparing our formulation (Proposed) with the standard-DAG formulation (Std-DAG),
we observed that Proposed had fewer numbers of additional gates.
We certainly succeeded in reducing the numbers of additional SWAP and Bridge gates
from standard-DAG formulation, which
is slightly better than fixed-layer formulation (Fixed-layer) as expected, for each of the coupling architectures.
By comparing Proposed with the original formulation without Bridge gates (No-Bridge),
we can see that Bridge gates indeed improve optimality of the formulation.
It is not surprising that the consideration of Bridge gates works best in LNN coupling architecture, which have leaves in the coupling graph.
Unfortunately,
the runtime of the exact algorithm grows exponentially with the number of qubits and gates;
while all runs for 5-qubit instances are within 1 minute, but those for 6-qubit instances are around 12 minutes.

\subsection{Comparison with other heuristic algorithms for larger circuits} \label{sec:exp-heuristic}
In the second set of experiments, we evaluated the heuristic algorithm
against two state-of-the-art heuristic algorithms. One is a randomized heuristic algorithm (called QRAND) implemented in Qiskit,
which is a Python software development kit for quantum programming~\cite{qiskit}.
The other is an $\mathrm{A}^*$-based heuristic search algorithm (called ZPW) proposed by Zulehner et al.~\cite{zulehner2018efficient}.
Their C++ implementation is available to the public, including their test circuit data, which originated from the RevLib benchmarks~\cite{soeken2011revkit}.
From their data, we chose 44 circuits with \#qubits ($|B|$) $\geq 10$ and \#gates $\leq 50,000$ for the experiment.
For all the circuits, we computed the mappings to the IBM Q 16 Rueschlikon V1.0.0 (ibmqx3) coupling architecture~\cite{ibmqx3}.
We implemented our algorithm on top of Qiskit v0.6. We set the parameter $\alpha=0.5$ and
restricted the $g$ of $d(K, g)$ to the gates within 10 steps from $K$.

\begin{table*}
	\centering
	\caption{Comparison of numbers of additional SWAP and Bridge gates in mappings with our proposed heuristic algorithm (Proposed),
  its original one without Bridge gates (No-Bridge), as well as QRAND and ZPW for circuits with 10 to 16 qubits from RevLib benchmark.
  The compositions of the numbers of SWAP and Bridge gates of Proposed are listed in the (\#SWAP+\#Bridge) column.
The runtime of Proposed are listed in the Time column. ($|B|$: Number of qubits used in circuit)}
	\label{tab:vs-others}
\small
	\begin{tabular}{lrr|rrrrlr} \hline
		Circuit name & $|B|$ & \#gates & QRAND & ZPW & No-Bridge & Proposed & (\#SWAP+\#Bridge) & Time [s] \\ \hline
mini\_alu\_305	 & 	10	 & 	173 	 & 	80 	 & 	46 	 & 	\textbf{40} 	 & 	41 & (28+13) & 0.5	 \\
qft\_10	 & 	10	 & 	200 	 & 	82 	 & 	40 	 & 	\textbf{33} 	 & 	\textbf{33} & (33+0)  & 1.9	 \\
sys6-v0\_111	 & 	10	 & 	215 	 & 	116 	 & 	67 	 & 	46 	 & 	\textbf{34} & (22+12)  & 0.6	 \\
rd73\_140	 & 	10	 & 	230 	 & 	100 	 & 	58 	 & 	49 	 & 	\textbf{34} & (23+11) & 0.6	 \\
ising\_model\_10	 & 	10	 & 	480 	 & 	18 	 & 	14 	 & 	\textbf{12} 	 & 	\textbf{12} & (12+0) & 0.7 	 \\
rd73\_252	 & 	10	 & 	5,321 	 & 	2,054 	 & 	1,541 	 & 	1,212 	 & 	\textbf{999} & (644+355)  & 20.1	 \\
sqn\_258	 & 	10	 & 	10,223 	 & 	4,060 	 & 	2,867 	 & 	2,254 	 & 	\textbf{1,875} & (1,108+767) & 57.5	 \\
sym9\_148	 & 	10	 & 	21,504 	 & 	8,001 	 & 	5,907 	 & 	4,456 	 & 	\textbf{3,770} & (1,856+1,914) & 237.5 	 \\
max46\_240	 & 	10	 & 	27,126 	 & 	10,833 	 & 	8,012 	 & 	5,905 	 & 	\textbf{4,932} & (2,573+2,359)  & 469.8	 \\
wim\_266	 & 	11	 & 	986 	 & 	385 	 & 	269 	 & 	225 	 & 	\textbf{155} & (65+90) & 4.8	 \\
dc1\_220	 & 	11	 & 	1,914 	 & 	721 	 & 	548 	 & 	446 	 & 	\textbf{329} & (154+175) & 6.4	 \\
z4\_268	 & 	11	 & 	3,073 	 & 	1,200 	 & 	907 	 & 	718 	 & 	\textbf{600} & (364+236)  & 11.2	 \\
life\_238	 & 	11	 & 	22,445 	 & 	9,181 	 & 	7,209 	 & 	5,264 	 & 	\textbf{4,539} & (2,638+1,901)  & 268.8 \\
9symml\_195	 & 	11	 & 	34,881 	 & 	14,470 	 & 	10,682 	 & 	8,124 	 & 	\textbf{6,630} & (3,909+2,721)  & 684.1	 \\
sym9\_146	 & 	12	 & 	328 	 & 	173 	 & 	85 	 & 	66 	 & 	\textbf{51} & (33+18)  & 0.9	 \\
cm152a\_212	 & 	12	 & 	1,221 	 & 	423 	 & 	341 	 & 	269 	 & 	\textbf{175} & (76+99)  & 4.3	 \\
sqrt8\_260	 & 	12	 & 	3,009 	 & 	1,263 	 & 	956 	 & 	728 	 & 	\textbf{587} & (377+210)  & 11.8	 \\
cycle10\_2\_110	 & 	12	 & 	6,050 	 & 	2,701 	 & 	1,856 	 & 	1,518 	 & 	\textbf{1,192} & (725+467)  & 27.9	 \\
rd84\_253	 & 	12	 & 	13,658 	 & 	5,817 	 & 	4,271 	 & 	3,271 	 & 	\textbf{2,784} & (1,777+1,007)  & 108.9	 \\
rd53\_311	 & 	13	 & 	275 	 & 	162 	 & 	98 	 & 	77 	 & 	\textbf{68} & (53+15) & 1.9 	 \\
ising\_model\_13	 & 	13	 & 	633 	 & 	28 	 & 	28 	 & 	\textbf{18} 	 & 	\textbf{18} & (18+0)  & 1.1	 \\
squar5\_261	 & 	13	 & 	1,993 	 & 	797 	 & 	605 	 & 	\textbf{422} 	 & 	446 & (285+161)  & 6.0	 \\
radd\_250	 & 	13	 & 	3,213 	 & 	1,347 	 & 	951 	 & 	721 	 & 	\textbf{623} & (382+241)  & 11.3	 \\
adr4\_197	 & 	13	 & 	3,439 	 & 	1,529 	 & 	1,044 	 & 	806 	 & 	\textbf{675} & (422+253)  & 12.6	 \\
root\_255	 & 	13	 & 	17,159 	 & 	7,172 	 & 	5,417 	 & 	4,070 	 & 	\textbf{3,516} & (2,160+1,356) & 152.4 	 \\
dist\_223	 & 	13	 & 	38,046 	 & 	16,548 	 & 	11,930 	 & 	9,326 	 & 	\textbf{7,879} & (4,952+2,927)  & 809.2	 \\
0410184\_169	 & 	14	 & 	211 	 & 	88 	 & 	76 	 & 	\textbf{44} 	 & 	45 & (35+10)  & 0.6	 \\
sym6\_316	 & 	14	 & 	270 	 & 	123 	 & 	70 	 & 	65 	 & 	\textbf{58} & (41+17) & 0.8 	 \\
cm42a\_207	 & 	14	 & 	1,776 	 & 	677 	 & 	494 	 & 	402 	 & 	\textbf{321} & (207+114) & 8.9 	 \\
cm85a\_209	 & 	14	 & 	11,414 	 & 	4,906 	 & 	3,598 	 & 	2,694 	 & 	\textbf{2,370} & (1,395+975)  & 75.8	 \\
clip\_206	 & 	14	 & 	33,827 	 & 	14,845 	 & 	11,011 	 & 	8,187 	 & 	\textbf{6,855} & (4,185+2,670) & 624.1 	 \\
sao2\_257	 & 	14	 & 	38,577 	 & 	16,974 	 & 	12,511 	 & 	9,188 	 & 	\textbf{7,479} & (4,691+2,788)  & 780.2	 \\
rd84\_142	 & 	15	 & 	343 	 & 	192 	 & 	103 	 & 	67 	 & 	\textbf{56} & (39+17) & 1.0 	 \\
misex1\_241	 & 	15	 & 	4,813 	 & 	1,844 	 & 	1,520 	 & 	1,216 	 & 	\textbf{1,067} & (602+465) & 21.8 	 \\
square\_root\_7	 & 	15	 & 	7,630 	 & 	3,243 	 & 	2,369 	 & 	\textbf{1,504} 	 & 	1,546 & (1,393+153)  & 46.7	 \\
ham15\_107	 & 	15	 & 	8,763 	 & 	3,635 	 & 	2,552 	 & 	1,979 	 & 	\textbf{1,685} & (980+705) 	 & 50.0 \\
dc2\_222	 & 	15	 & 	9,462 	 & 	4,112 	 & 	2,933 	 & 	2,326 	 & 	\textbf{1,798} & (1,099+699)  & 58.5	 \\
co14\_215	 & 	15	 & 	17,936 	 & 	8,423 	 & 	6,566 	 & 	4,318 	 & 	\textbf{3,294} & (1,780+1,514)  & 183.9	 \\
cnt3-5\_179	 & 	16	 & 	175 	 & 	69 	 & 	54 	 & 	38 	 & 	\textbf{35} & (23+12)  & 0.5	 \\
cnt3-5\_180	 & 	16	 & 	485 	 & 	183 	 & 	124 	 & 	98 	 & 	\textbf{88} & (49+39) 	 & 3.0 \\
qft\_16	 & 	16	 & 	512 	 & 	296 	 & 	117 	 & 	\textbf{82} 	 & 	\textbf{82} & (82+0)  & 4.8	 \\
ising\_model\_16	 & 	16	 & 	786 	 & 	20 	 & 	24 	 & 	\textbf{18} 	 & 	\textbf{18} & (18+0)  & 1.3	 \\
inc\_237	 & 	16	 & 	10,619 	 & 	4,351 	 & 	3,138 	 & 	2,542 	 & 	\textbf{2,098} & (1,198+900)  & 68.3	 \\
mlp4\_245	 & 	16	 & 	18,852 	 & 	8,104 	 & 	6,212 	 & 	4,547 	 & 	\textbf{3,988} & (2,495+1,493) & 179.7 	 \\ \hline
	\end{tabular}
\end{table*}

The proposed heuristic algorithm outperformed QRAND and ZPW for all instances.
See Table~\ref{tab:vs-others} for all results.
The numbers of additional SWAP and Bridge gates in the mappings with the heuristic algorithm (Proposed) decreased by
10.0--72.3\% (Avg. 53.2\%) and 10.9--48.8\% (Avg. 34.7\%)
from those of QRAND and ZPW, respectively.
Even without Bridge gates (No-Bridge), in the mappings with commutation rules
the numbers of additional SWAP gates decreased by 10.0--72.3\% (Avg. 45.5\%) and 7.1--42.1\% (Avg. 23.8\%)
from those of QRAND and ZPW, respectively.
From the columns of Proposed and No-Bridge, we can confirm that
Bridge gates can find better mappings in almost all cases except for
a few small circuits. The numbers of additional gates decreased by 14.2\% on average.
This implies, in spite of larger search space due to the consideration of Bridge gates,
the proposed heuristic algorithm can efficiently explore it to find better solutions.
All runs of heuristic algorithms were less than 15 minutes.

Since we needed to give the initial layout $l_0$ for our algorithm and QRAND,
we used the trivial layout $l_0: (b_1,\dots,b_{|Q|}) \mapsto (q_1,\dots,q_{|Q|})$.
The number of additional gates may possibly be further reduced by finding a good initial layout for the heuristics.
For our algorithm, we added a post-processing to remove leading useless
SWAP gates by changing the given initial layout. The reverse
traversal technique by Li et al.~\cite{li2019tackling} can be applied instead of the post-processing,
and to find a better initial layout.

\section{Conclusion}
We addressed the problem of mapping quantum circuits to NISQ computers, whose operations are limited by their coupling architecture,
with as few additional gates as possible. Our proposed solution to the mappings is to use gate commutation rules and gate transformation
rules in the form of SWAP and Bridge gates. Many previous studies did not consider such gate commutation rules or Bridge gates
to reduce additional gates in the mappings. We developed exact and heuristic algorithms that take advantage of such rules. Comparing them with
the state-of-the-art circuit mapping algorithms, we demonstrated that our proposed algorithms
can find better mappings with fewer additional gates for the circuits of a commonly used benchmark dataset.

\small
\bibliographystyle{elsarticle-num}
\bibliography{qc-opt}

\appendix
\section{Construction of dependency graph} \label{sec:const-dg}

We give the algorithm for constructing dependency graph in Algorithm~\ref{algo:dependency}.
Because it is specifically tailored to the commutation rules shown in Fig.~\ref{fig:rules},
the commutation rules are not stated as input to the algorithm.

\begin{algorithm}
\caption{Constructing dependency graph}
\label{algo:dependency}
\begin{algorithmic}[1]
\REQUIRE List of gates $L$ in a given logical circuit
\ENSURE Dependency graph $D$
\STATE $V_D \leftarrow \{g \mid g \in L\}$\quad\COMMENT{node set of $D$}
\STATE $E_D \leftarrow \emptyset$\quad\COMMENT{edge set of $D$}
\FORALL{gate pair $(g_i, g_j)$ such that $i < j$ in $L$} \label{line:dep-for}
  \FORALL{common acting qubit $b$ of $g_i$ and $g_j$}
    \STATE $S \leftarrow \{s \mid$ symbol $s$ between $g_i$ and $g_j$ on $b\}$
    \IF{$S \not\subseteq \{R_z, \bullet\}$ and $S \not\subseteq \{R_x, \oplus\}$}
      \STATE $E_D \leftarrow E_D \cup \{(g_i, g_j)\}$
    \ENDIF
  \ENDFOR
\ENDFOR
\STATE \textbf{return} $(V_D, E_D)$
\end{algorithmic}
\end{algorithm}

Note that the dependency graph obtained by the algorithm is redundant.
If necessary, the minimal set of edges can be obtained by checking
each of the edges in $E_D$ and removing the edge $(s,t)$
if there exists a path from $s$ to $t$
in the graph with reduced edge set excluding $(s,t)$.

\end{document}